\documentclass[twocolumn,floatfix,prb,aps,showpacs]{revtex4-1}
\usepackage{graphicx,amsmath,amssymb,color}
\usepackage{nicefrac,subfigure}
\usepackage[titletoc,title]{appendix}
\usepackage{pbox}

\newcommand{\be}{\begin{equation}}
\newcommand{\ee}{\end{equation}}

\newcommand{\ba}{\begin{eqnarray}}
\newcommand{\ea}{\end{eqnarray}}

\begin{document}

\title{Positions of the magnetoroton minima in the fractional quantum Hall effect}
\author{Ajit C. Balram$^{1,2}$ and Songyang Pu$^2$}
\affiliation{
   $^{1}$Niels Bohr International Academy and the Center for Quantum Devices,
Niels Bohr Institute, University of Copenhagen, 2100 Copenhagen, Denmark}
\affiliation{
   $^{2}$Department of Physics, 104 Davey Lab, Pennsylvania State University, University Park, Pennsylvania 16802, USA}

\begin{abstract}
\pacs{73.43.-f, 71.10.Pm}
The multitude of excitations of the fractional quantum Hall state are very accurately understood, microscopically, as excitations of composite fermions across their Landau-like $\Lambda$ levels. In particular, the dispersion of the composite fermion exciton, which is the lowest energy spin conserving neutral excitation, displays filling-factor-specific minima called ``magnetoroton" minima. Simon and Halperin employed the Chern-Simons field theory of composite fermions [Phys. Rev. B {\bf 48}, 17368 (1993)] to  predict the magnetoroton minima positions. Recently, Golkar \emph{et al.} [Phys. Rev. Lett. {\bf 117}, 216403 (2016)] have modeled the neutral excitations as deformations of the composite fermion Fermi sea, which results in a prediction for the positions of the magnetoroton minima.  Using methods of the microscopic composite fermion theory we calculate the positions of the roton minima for filling factors up to 5/11 along the sequence $s/(2s+1)$ and find them to be in reasonably good agreement with both the Chern-Simons field theory of composite fermions and Golkar \emph{et al.}'s theory. We also find that the positions of the roton minima are insensitive to the microscopic interaction in agreement with Golkar \emph{et al.}'s theory. As a byproduct of our calculations, we obtain the charge and neutral gaps for the fully spin polarized states along the sequence $s/(2s\pm 1)$ in the lowest Landau level and the $n=1$ Landau level of graphene.
\end{abstract}
\maketitle

\section{Introduction}
When electrons are confined to two dimensions and subjected to a strong perpendicular magnetic field, they exhibit the marvelous phenomena of the fractional quantum Hall effect (FQHE) \cite{Tsui82}. FQHE arises due to the formation of topological particles called composite fermions \cite{Jain89,Jain07} (CFs), which are bound states of an electron and an even number ($2p$) of quantized vortices. Due to the binding of vortices CFs feel a reduced magnetic field compared to the external magnetic field. In this reduced magnetic field CFs form their own Landau-like levels, called $\Lambda$ levels ($\Lambda$Ls). The electron filling factor $\nu$ is related to the filling factor of composite fermions $\nu^{*}$ as $\nu=\nu^{*}/(2p\nu^{*}\pm1)$. The $\nu^*=s$ integer quantum Hall effect (IQHE) of composite fermions produces incompressible states at the Jain fractions $\nu=s/(2ps\pm 1)$, which explains a vast majority of the FQHE phenomena occurring in the lowest Landau level (LLL). \\

The neutral excitations of the FQHE were first considered by Girvin, MacDonald and Platzmann \cite{Girvin85,Girvin86} through a single-mode approximation (SMA), which treats the neutral excitations as LLL projected density waves. The SMA was shown to give a good description of the neutral excitation at $1/3$, in particular uncovering a minimum in the dispersion termed the ``magnetoroton minimum'', but was subsequently found not to work well for other FQH states \cite{Jain97}. Further progress in our understanding of the neutral excitations became possible with the microscopic CF theory, where the lowest energy neutral mode was naturally understood as an exciton (i.e. a particle hole pair) of composite fermions, obtained by exciting a composite fermion from the topmost occupied $\Lambda$ level to the lowest unoccupied $\Lambda$ level. The dispersions of the neutral mode, and in particular their roton minima, were obtained at the Jain fractions $s/(2ps\pm 1)$ from the microscopic theory; they are in excellent quantitative agreement with exact diagonalization studies \cite{Dev92,Jain97,Kang01,Peterson03,Moller05,Majumder09,Rhone11,Yang12b,Yang13a,Balram13,Yang14,Jain14,Majumder14} and in good qualitative and semi-quantitative agreement with experiments \cite{Pinczuk93,Davies97,Scarola00,Kang01,Groshaus08,Kukushkin09,Rhone11}.  \\

Effective field theories have also been used to address this issue \cite{Zhang92}. Notably, the Chern-Simons theory of composite fermions \cite{Lopez91,Halperin93} has been employed to calculate the dispersion of the neutral modes \cite{Lopez93,Simon93,Wang17}. Employing these methods with a modification of the Random Phase Approximation (RPA), Simon and Halperin \cite{Simon93} predicted that the magnetoroton minima at $\nu=s/(2s+1)$ occur at:
\begin{equation}
k\ell^{\rm CS}\approx\frac{1}{\sqrt{2s(2s+1)}} \Big(i+\frac{1}{4}\Big) \pi
\label{Simon_Halperin_pred_value}
\end{equation}
where $i$ is an integer. The approximations used in obtaining the above result are expected to work well in the long wavelength limit and at large values of $s$ (at filling factors close to 1/2). \\

Recently, Golkar \emph{et al.} \cite{Golkar16} treated the neutral excitations as quantized shape deformations of the composite fermion Fermi surface at 1/2. From this approach, they predict that the magnetoroton minima at $\nu=s/(2s+1)$ are located at:
\begin{equation}
k\ell=\frac{x_{i}}{2s+1} 
\label{Golkar_pred_value}
\end{equation}
where $x_{i}$ are the zeros of the Bessel function of order one of the first kind $J_{1}(x)$ \cite{Davis27}, $k$ is the magnitude of the planar wave vector and $\ell=\sqrt{\hbar/(eB)}$ is the magnetic length. These values are predicted to be insensitive to the specific microscopic details of the Hamiltonian. The difference between successive roots of $J_{1}(x)$ is approximately constant and approaches $\pi$ asymptotically \cite{Davis27}. Thus the successive magnetoroton minima positions are expected to be approximately equidistant from each other. We emphasize here that Golkar \emph{et al.}'s theory is based on the assumption that the coupling between the composite fermions and the dynamical gauge field is infinitely strong, which becomes better and better in the limit $\nu \rightarrow 1/2$. In addition, their derivation mandates that $k\ell \ll 1$ and hence their predictions are expected to be more accurate for the first few roton minima. \\

The positions of magnetoroton minima predicted by Eq. \ref{Simon_Halperin_pred_value} at $\nu=1/3$ agree with those predicted by Eq. \ref{Golkar_pred_value} within $30\%$; for 3/7 the deviation is $10\%$; and for the states $s/(2s+1)$ with $s>9$ the difference between the two predictions is less than $5\%$. This is understood by noting that $x_{i}\approx (i+1/4)\pi$ (this approximation improves with increasing $i$ and is asymptotically exact) \cite{Davis27}, and thus for large $s$ the two predictions are approximately equal (see Table \ref{tab_minima_positions magnetoroton}). \\

Can we use the SMA to test these predictions for the positions of the magnetoroton minima? The SMA is not well-suited to test either Simon and Halperin's theory or Golkar \emph{et al.}'s theory. Firstly, the SMA predicts only $s$ magneto-roton minima for states along the sequence $s/(2s+1)$\cite{Park00a}. Secondly, in the sequence of states $s/(2s+1)$ only the excitations of 1/3 is well-described by the SMA\cite{Girvin85,Girvin86,Jain97}. On the other hand, along the sequence $s/(2s+1)$, 1/3 is the furthest from 1/2, so one would expect the above predictions to be least accurate for this state. Thus, we have to resort to the well-established microscopic composite fermion theory to accurately test these predictions (since most of these states are beyond the reach of exact diagonalization). \\

In the limit of zero LL mixing and electrons interacting via a two-body interaction, there is an exact particle-hole symmetry in the LLL \cite{Girvin84,Son15,Balram15b,Balram16b}, which ascertains that the postitions of the magnetoroton minma at $\nu=s/(2s+1)$ and $\nu=1-s/(2s+1)=(s+1)/(2s+1)$ coincide with each other. In this paper we use the microscopic formulation of the composite fermion theory, which obeys particle-hole symmetry to a good extent \cite{Balram16b}, to calculate the positions of the magnetoroton minima for filling factors up to 5/11 along the sequence $s/(2s+1)$. (Due to technical reasons we shall only consider states along the sequence $s/(2s+1)$ and conclusions drawn from them can be extended to states at $\nu=(s+1)/(2s+1)$ using the aforementioned particle-hole symmetry.) Our calculations show that the predictions of both Simon and Halperin and Golkar \emph{et al.} are reasonably accurate for the roton minima. For large values of $s$ (which is the regime where Golkar \emph{et al.}'s and Simon and Halperin's theories are expected to work well), Golkar \emph{et al.}'s and Simon and Halperin's numbers agree well with each other and with the microscopic CF theory. Furthermore, as predicted by Golkar \emph{et al.}, we find that these magnetoroton minima positions are nearly independent of the specific details of the microscopic Hamiltonian. For completeness, we also extend previous calculations \cite{Jain97,Park00,Morf02,Jain07} of the charge and neutral gaps to further Jain FQH states in the LLL and the $n=1$ LL of graphene. \\

The insensitivity of the positions of the roton minima to the specific details of the microscopic interaction was also understood within the microscopic theory of composite fermions \cite{Balram15c}. Within the microscopic theory, this follows from the facts that (i) roton minima arise due to the complex density profiles of the constituent CF particle and CF hole, which produce especially low energies when the distance between them (which is related to the wave vector) is such that the maxima of one coincides with the minima of the other; and (ii) the wave functions are ``universal'' and hence insensitive to the interaction provided that the short range part dominates. Furthermore, Ref. \cite{Kamilla96b} showed that the number and positions of the strong minima are roughly the same for the FQHE state and the corresponding IQHE state to which it is related by the composite fermion theory. \\

The paper is organized as follows: In the next section we give an introduction to the spherical geometry and the microscopic composite fermion theory. In Sec. \ref{sec:results} we present our results on the CF exciton dispersion for FQH states along the sequence $s/(2s+1)$. Finally, in Sec. \ref{sec:conclusions} we conclude the paper with a summary of our results. \\

\section{Methods}
\label{sec:methods}
All our calculations are carried out in the spherical geometry \cite{Haldane83,Greiter11} in which the two-dimensional plane of electrons is wrapped on the surface of a sphere and a radial magnetic field of strength $2Q\phi_{0}$ ($\phi_{0}=hc/e$ is a quantum of flux) is generated by a Dirac monopole ($2Q$ is an integer) sitting at the center of the sphere. Appropriate to this geometry the total orbital angular momentum $L$ is a good quantum number and incompressible ground states have $L=0$. The magnitude of the wave vector in the planar geometry $k$ is related to $L$ as $k=L/R$, where $R=\sqrt{Q}\ell$ is the radius of the Haldane sphere \cite{Jain97,Jain07}. Throughout this work we shall restrict to the case of fully spin polarized electrons and hence the total spin angular momentum quantum number $S=N/2$. The effective flux experienced by composite fermions is $2Q^{*}=2Q-2p(N-1)$ and an incompressible FQH state with $s$ filled $\Lambda$Ls arises when $2Q^{*}=(N-s^{2})/s$. The ground state of the Coulomb interaction at $s/(2s+1)$ is very well described by the following Jain wave function \cite{Jain89,Jain90,Jain07}:
\begin{equation}
 \Psi_{s/(2s+1)}=\mathcal{P}_{\rm LLL}\prod_{j<k}(u_{j}v_{k}-u_{k}v_{j})^{2}\Phi_{s}
 \label{Jain_wf_gs}
\end{equation}
where $u=\cos(\theta/2)e^{i\phi/2}$ and $v=\sin(\theta/2)e^{-i\phi/2}$ are the spinor coordinates with $\theta$ and $\phi$ being the polar and azimuthal angles on the sphere respectively. $\Phi_{s}$ is the wave function of $s$ filled LLs of electrons and $\mathcal{P}_{\rm LLL}$ implements LLL projection which we carry out using the Jain-Kamilla method \cite{Jain97,Jain97b,Jain07}. \\

In analogy to the ground state wave function, the lowest energy CF exciton wave function at total orbital angular momentum $L$ is given by:
\begin{equation}
 \Psi^{L,\rm CF~exciton}_{s/(2s+1)}=\mathcal{P}_{\rm LLL}\prod_{j<k}(u_{j}v_{k}-u_{k}v_{j})^{2}\Phi_{s}^{L,\rm exciton}
 \label{Jain_wf_ex}
\end{equation}
where $\Phi_{s}^{L,\rm exciton}$ is the exciton wave function in the corresponding IQHE system with a hole in the otherwise full $(s-1)^{\rm th}$ LL and a particle in the otherwise empty $s^{\rm th}$ LL. The Jastrow factor $J=\prod_{j<k}(u_{j}v_{k}-u_{k}v_{j})$ has $L=0$ and projection to the LLL does not change $L$ \cite{Rezayi91,Jain07} and thus we use the same $L$ on the two sides of the above equation. By angular momentum addition of the constituent particle and hole, the exciton in the IQHE system has one state at each $L$ (with the standard $2L+1$ degeneracy arising from the $L_{z}$ values which we suppress) ranging from $L_{\rm min}=1$ to $L_{\rm max}=(Q^{*}+s-1)+(Q^{*}+s)$. It turns out that the $L=1$ exciton gets killed upon composite-fermionization (multiplication of an IQHE state by $J^2$ followed by projection to the LLL) \cite{Dev92,Balram13} and therefore the angular momentum of the exciton for the FQH state at $s/(2s+1)$ ranges from $L_{\rm min}=2$ to $L_{\rm max}=2Q^{*}+2s-1$. We use these wavefunctions to evaluate the energy for a real space interaction $V(r)$ using the Metropolis Monte Carlo method \cite{Binder10}. A more accurate scheme called CF diagonalization \cite{Mandal02,Jain07} can be used to improve upon these numbers but Majumder and Mandal \cite{Majumder14} have shown that for states along the sequence $s/(2s+1)$ just evaluating the CF exciton is sufficient to estimate the positions of the magnetoroton minima since CF diagonalization within the subspace of a few single excitons does not change these positions significantly. However, CF diagonalization may be necessary to study the CF exciton dispersion for FQH states described by composite fermions carrying more than two vortices, since here CF-$\Lambda$L mixing is known to alter the magnetoroton minima positions \cite{Peterson03}. \\

For $V(r)$ we consider the $1/r$ Coulomb interaction, which describes the physics of the LLL, in detail. Besides that, for $\nu=1/3,~2/5$ and $3/7$ we look at the effective interaction of Ref. \cite{Balram15c} which simulates the physics of the $n=1$ LL of graphene, whose FQH states are also well described by the microscopic CF theory. Ref. \cite{Balram15c} reported on the exciton dispersion in the LLL and the $n=1$ LL of graphene for a particular system at $\nu=2/5$. The positions of the magneto-roton minima for these two cases, obtained from very different interactions, coincide. Here, we follow up on their work to evaluate the exciton dispersion in detail at $\nu=1/3,~2/5$ and $3/7$ in the $n=1$ LL of graphene. To further investigate the robustness of the position of the magnetoroton minima to LL mixing we have studied the ``unprojected'' Jain wave functions \cite{Trivedi91,Kamilla97b} (i.e. Eq. \ref{Jain_wf_ex} without the LLL projection), which have a finite amplitude in the higher LLs, at $\nu=1/3,~2/5,~3/7$ and $4/9$. We note that the unprojected wave functions do not give a fully realistic account of LL mixing (such a treatment of LL mixing is beyond the scope of this work, see Ref. \cite{Zhang16} for one such recent attempt), but it is likely that they are adiabatically connected to the true Coulomb ground states (for $\nu=2/5$ Ref. \cite{Rezayi91} explicitly demonstrated that the unprojected state is adiabatically connected to the projected state, which in turn is very likely to be adiabatically connected to the actual Coulomb state). We use the $n=1$ LL of graphene and the unprojected state dispersions to serve as a test for the robustness of the positions of the magnetoroton minima to microscopic details.  \\

Furthermore, from the CF exciton dispersions we extract the charge and neutral Coulomb gaps of FQH states. The charge gap, which appears as the activation energy in transport experiments, is defined as the $k\ell\rightarrow \infty$ limit of the CF exciton dispersion. The neutral gap is the energy difference between the lowest lying CF exciton state and the ground state; it is the excitation energy of the global magnetoroton minimum and can be measured using phonon absorption \cite{Mellor95,Zeitler99} or inelastic light scattering \cite{Pinczuk93,Davies97,Pinczuk98,Groshaus08} experiments. For completeness we also estimate the long wavelength limit ($k\ell\rightarrow 0$) of the CF exciton gap. All the energy gaps shown below include the so-called ``density'' correction. This takes into account the fact that the density for a finite system in the spherical geometry depends on the number of electrons $N$ and differs slightly from its thermodynamic value. To mitigate this effect of the $N$-dependence we use the density corrected energy gap \cite{Morf86} $\Delta E_{N}'=\sqrt{2Q\nu/N}\Delta E_{N}$, which is then extrapolated to the thermodynamic limit $N\rightarrow \infty$. The error bars shown below are computed from the statistical uncertainity of the Monte Carlo sampling.   \\

\section{Results}
\label{sec:results}
\begin{figure*}[t]
\begin{center}
\includegraphics[width=7.5cm,height=5.0cm]{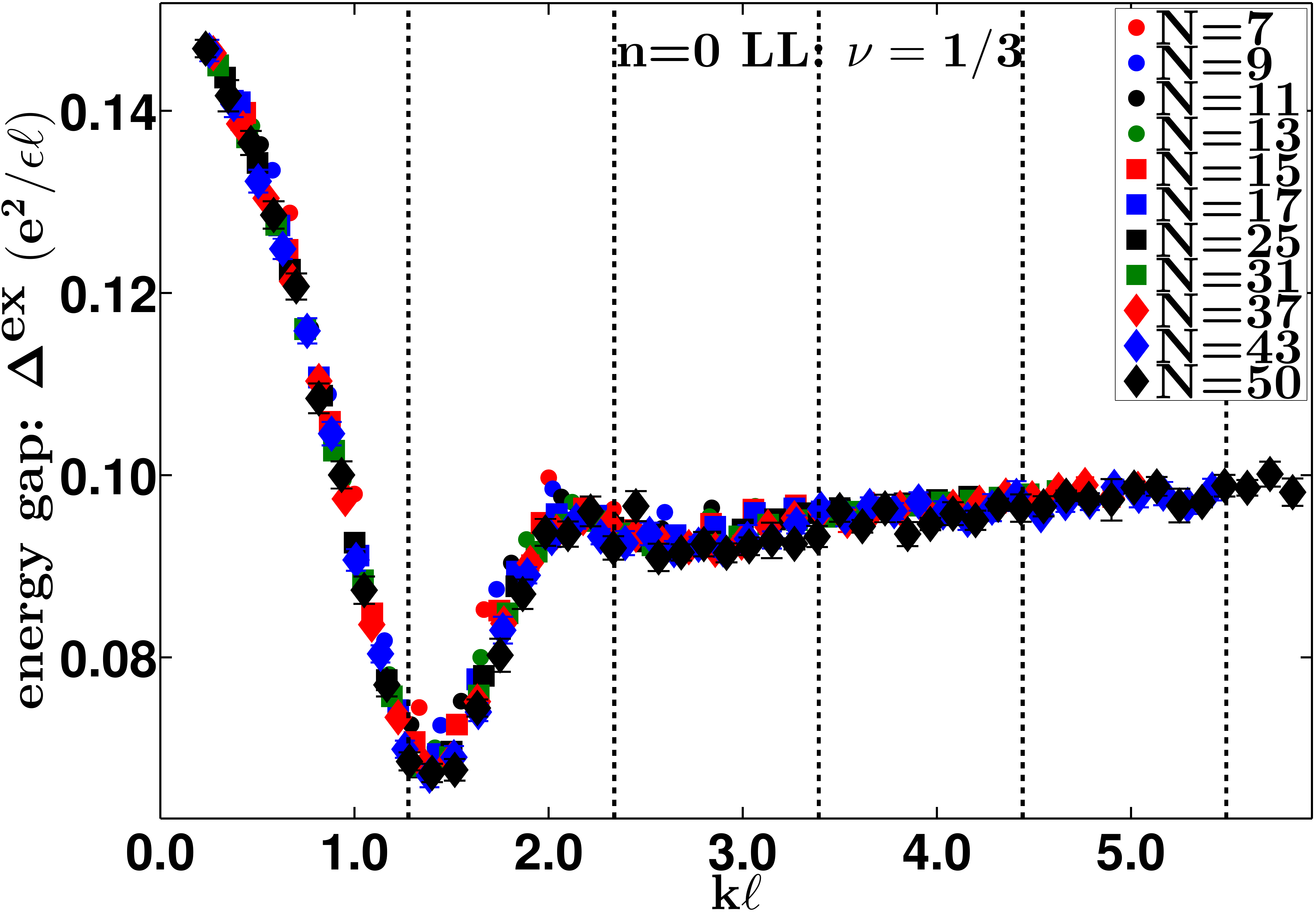}
\includegraphics[width=7.5cm,height=5.0cm]{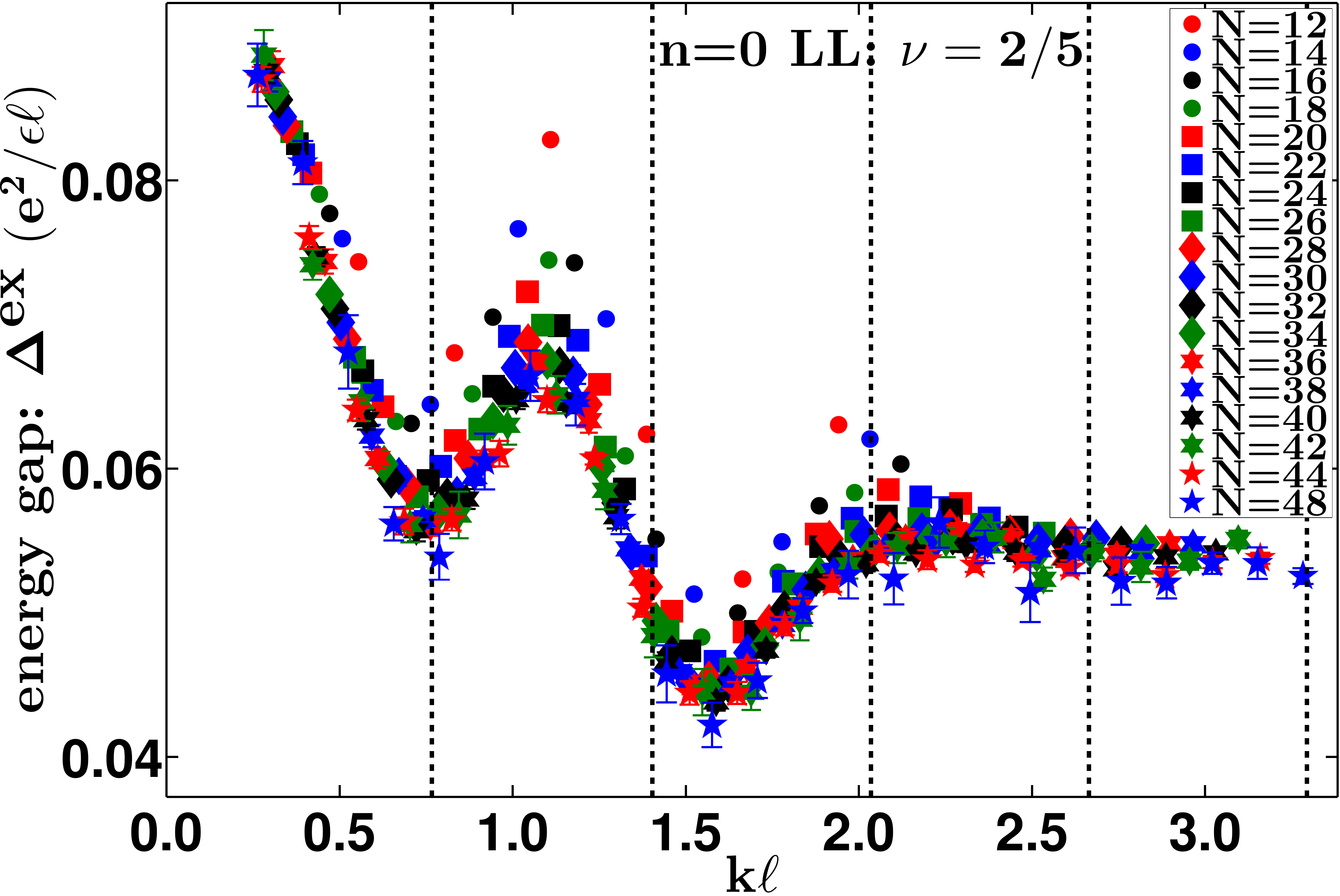} \\
\includegraphics[width=7.5cm,height=5.0cm]{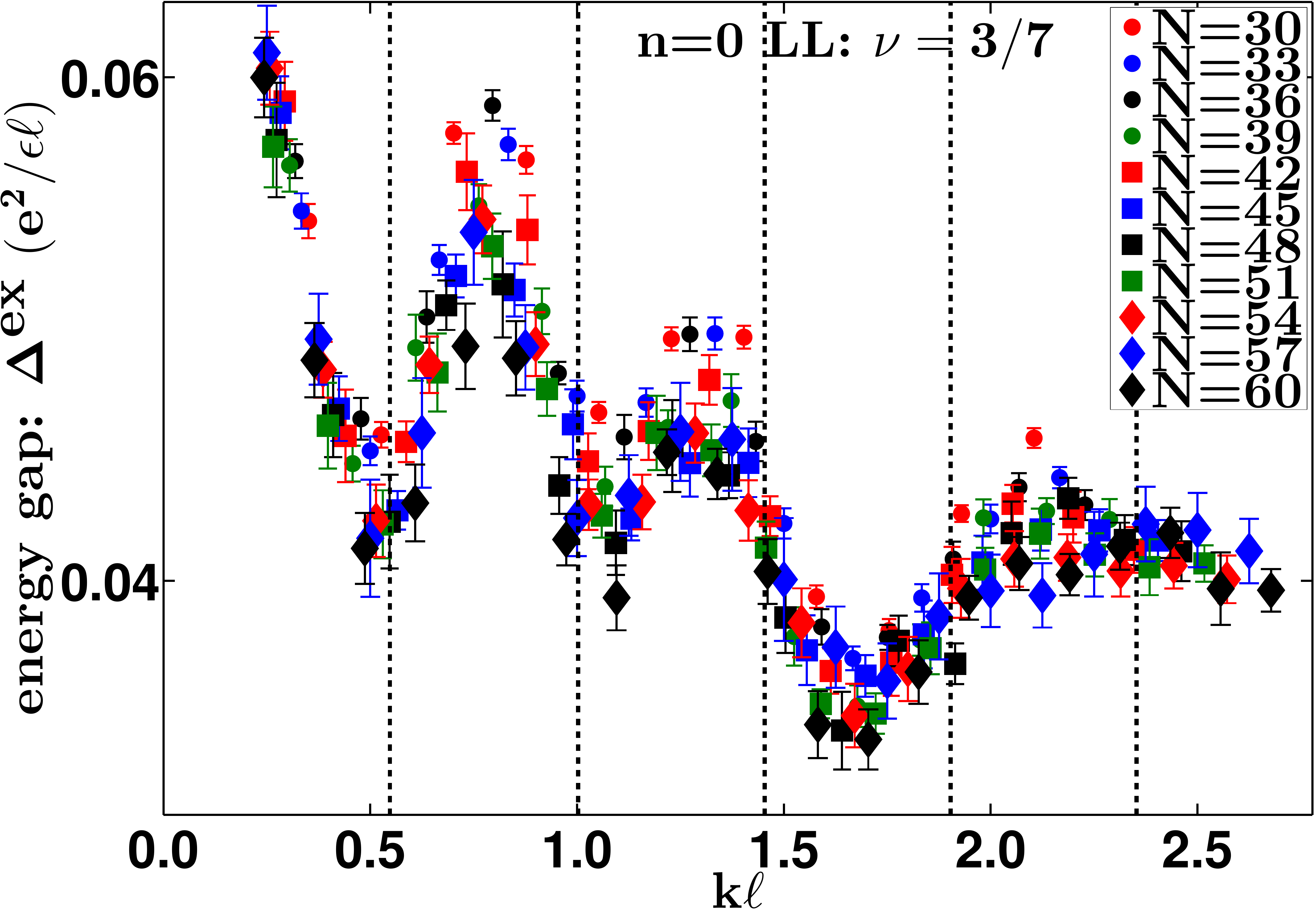}  
\includegraphics[width=7.5cm,height=5.0cm]{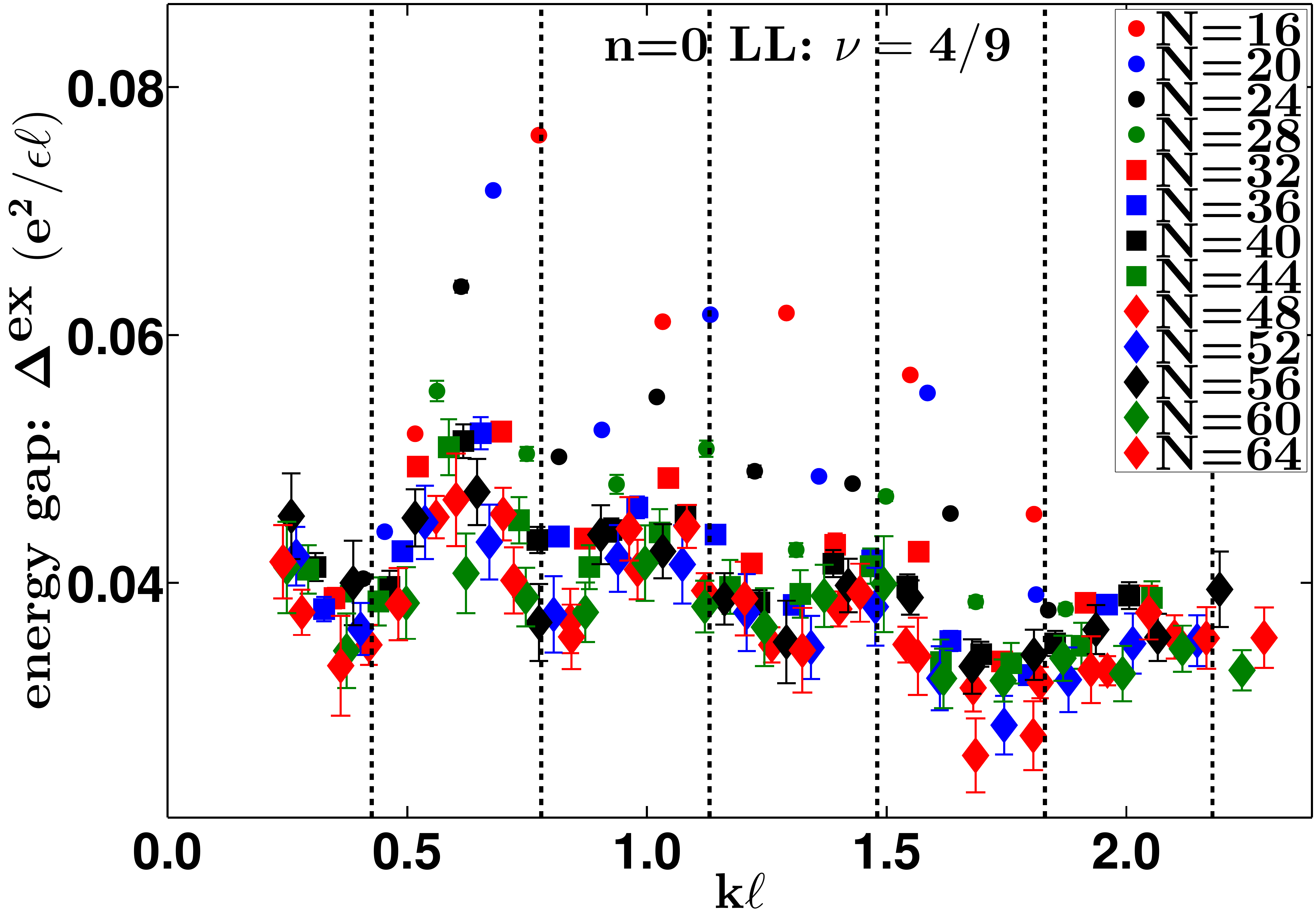} \\
\includegraphics[width=7.5cm,height=5.0cm]{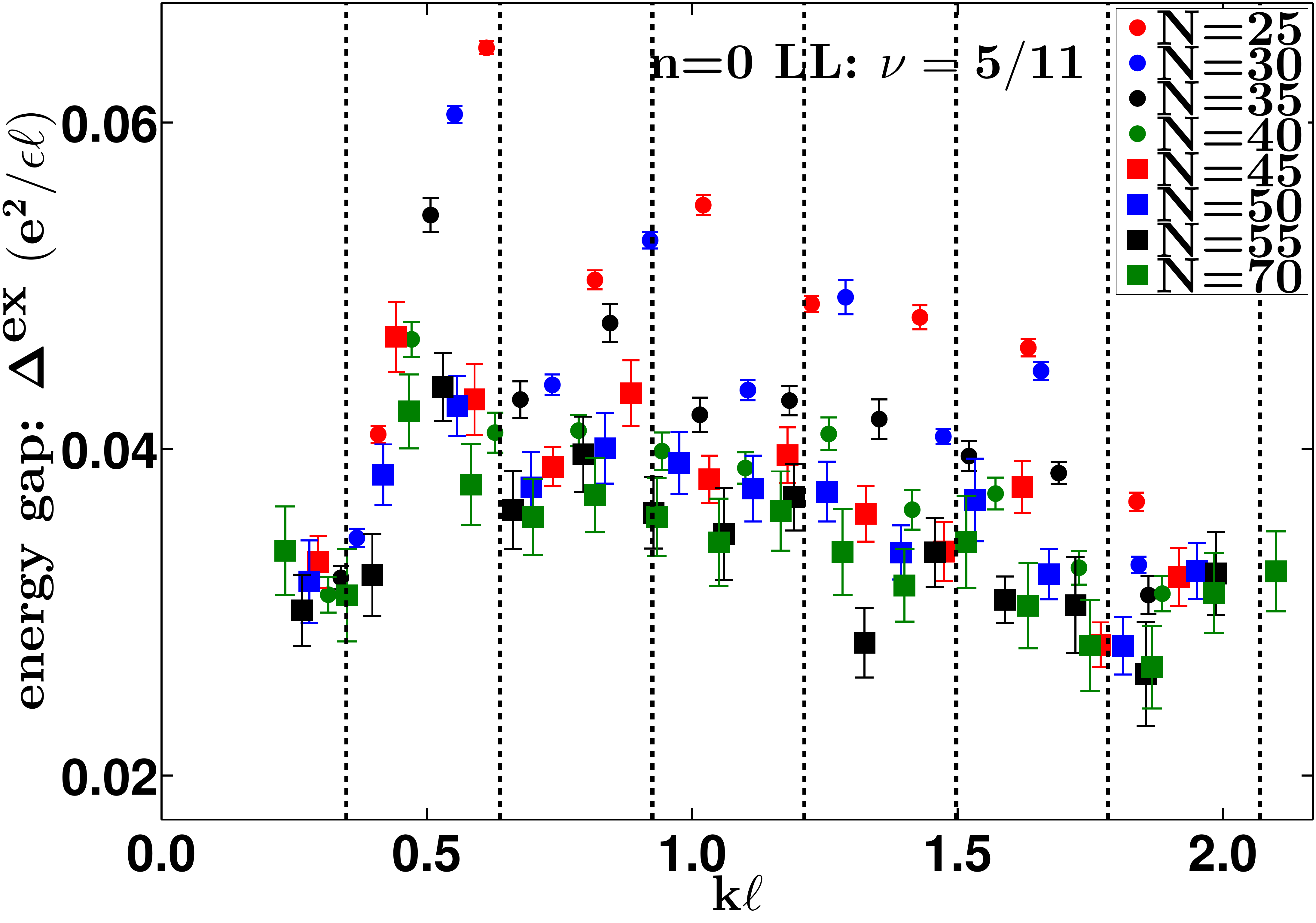}  
\end{center}
\caption{CF exciton dispersion in the LLL for fully spin polarized states at various filling factors along the sequence $s/(2s+1)$ obtained in the spherical geometry. The vertical black dashed lines show the values predicted by Golkar \emph{et al.} \cite{Golkar16}. Some of these exciton dispersions have been reproduced from Refs. \cite{Balram13,Balram15c}.}
\label{CF_exciton_dispersion}
\end{figure*}

\begin{figure*}[t]
\begin{center}
\includegraphics[width=7.5cm,height=5.0cm]{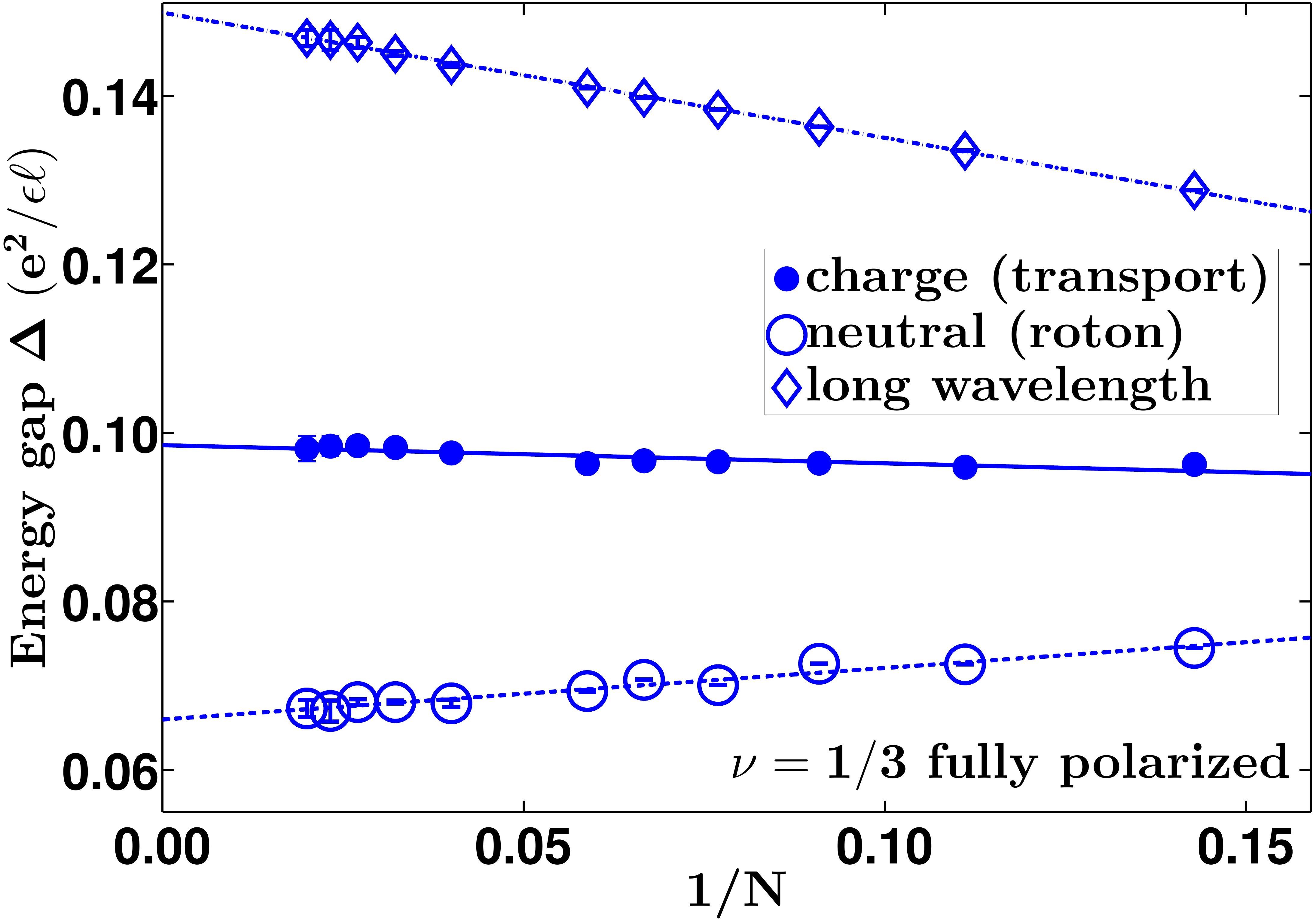}
\includegraphics[width=7.5cm,height=5.0cm]{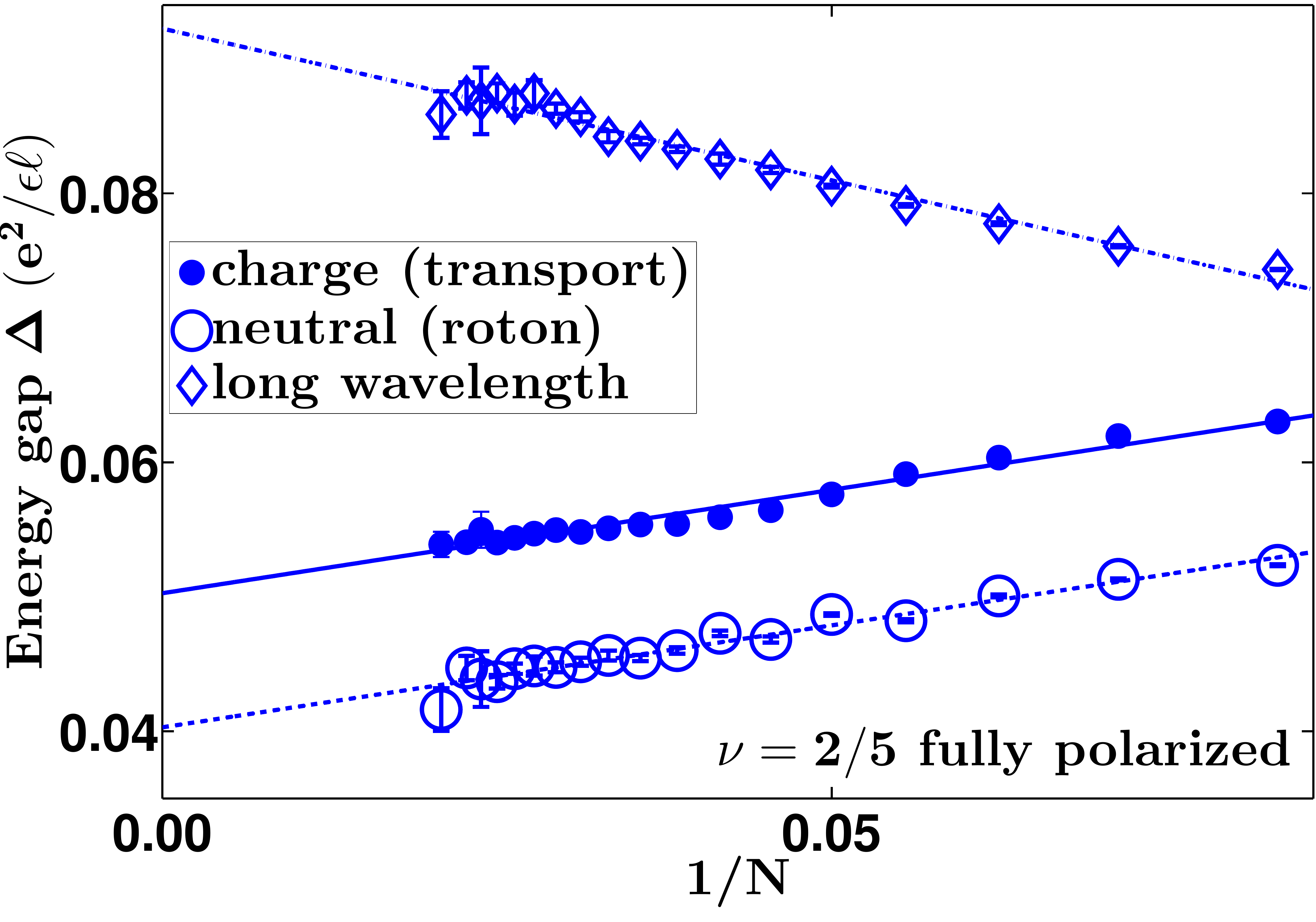} \\
\includegraphics[width=7.5cm,height=5.0cm]{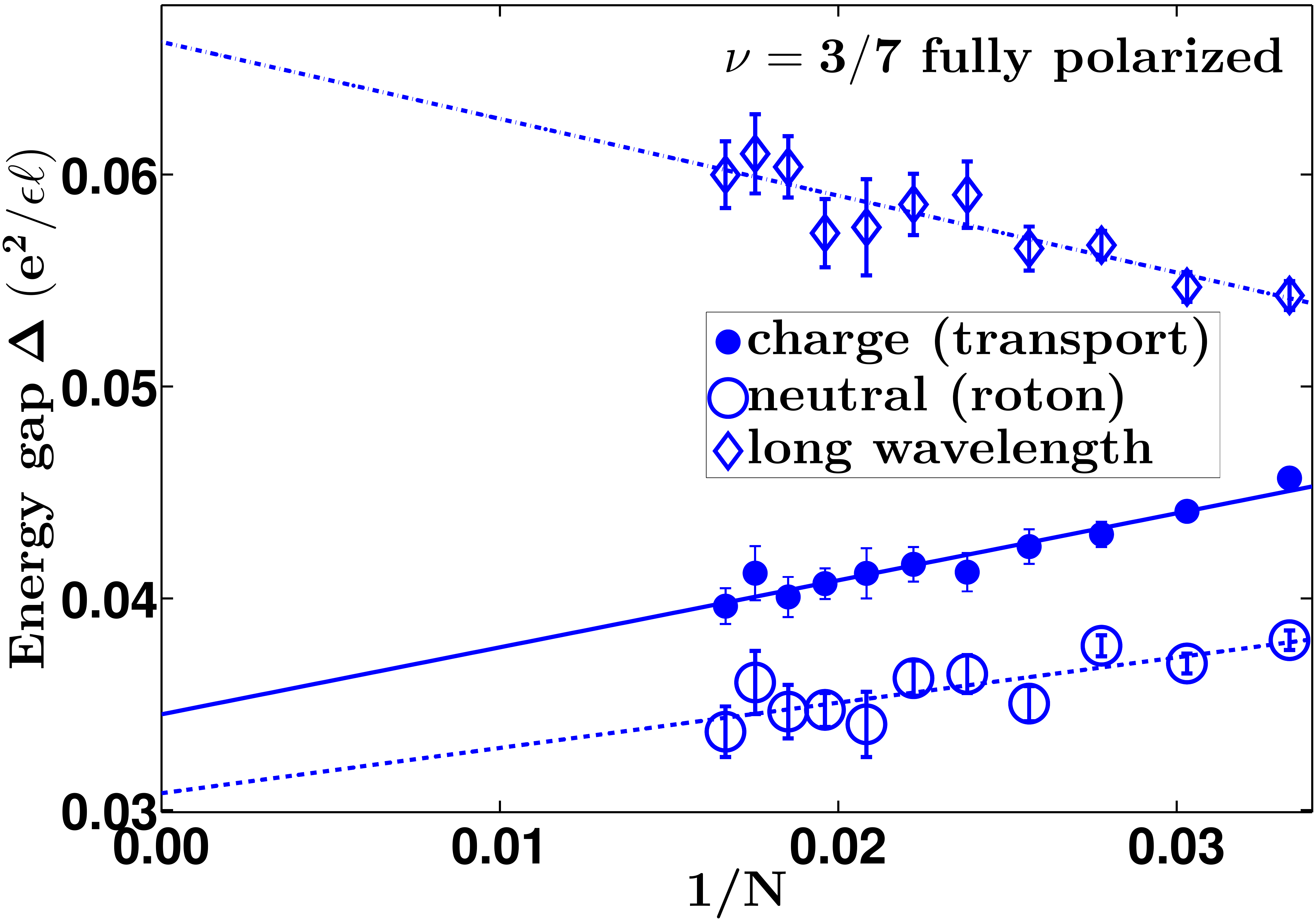}  
\includegraphics[width=7.5cm,height=5.0cm]{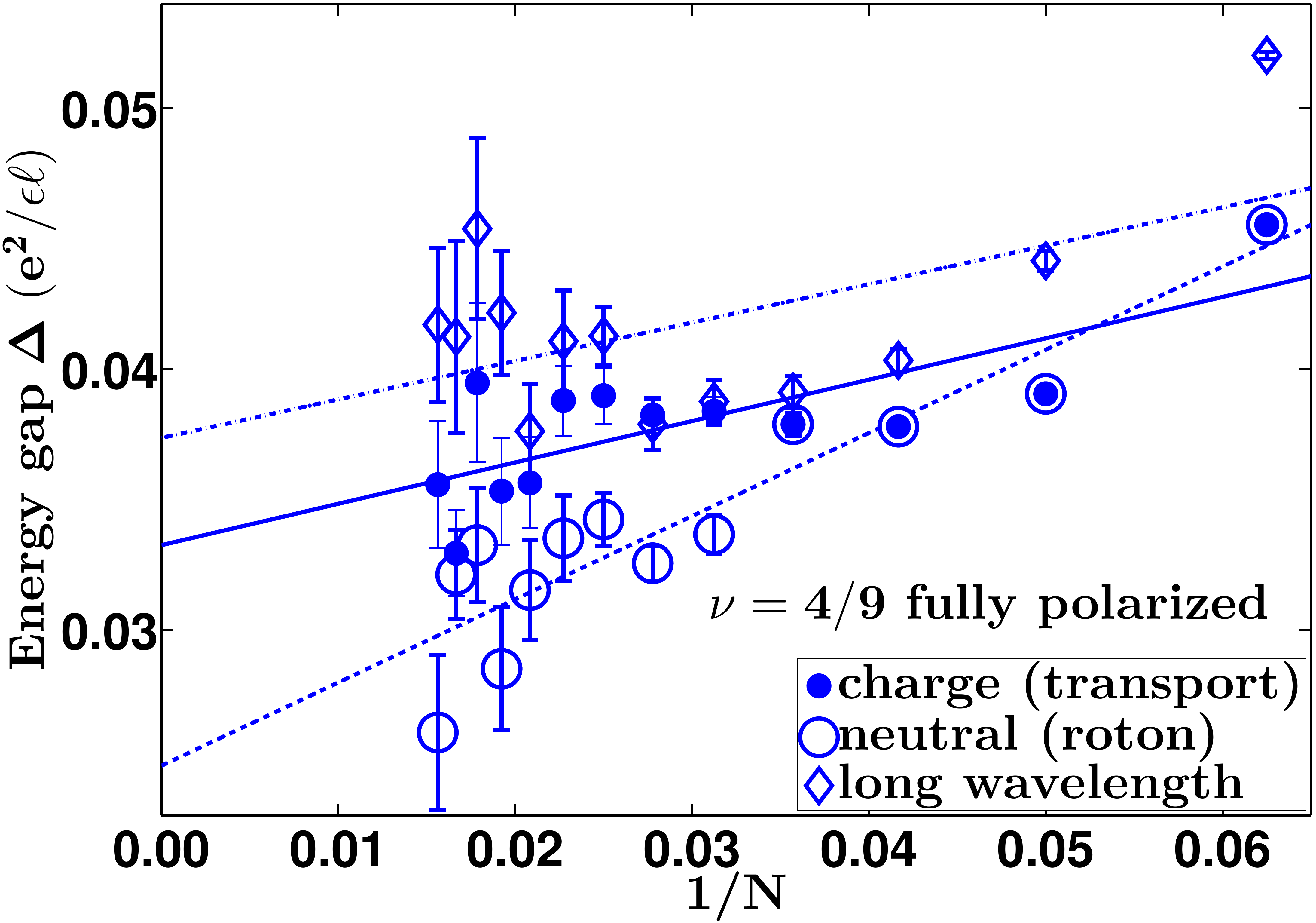} \\
\includegraphics[width=7.5cm,height=5.0cm]{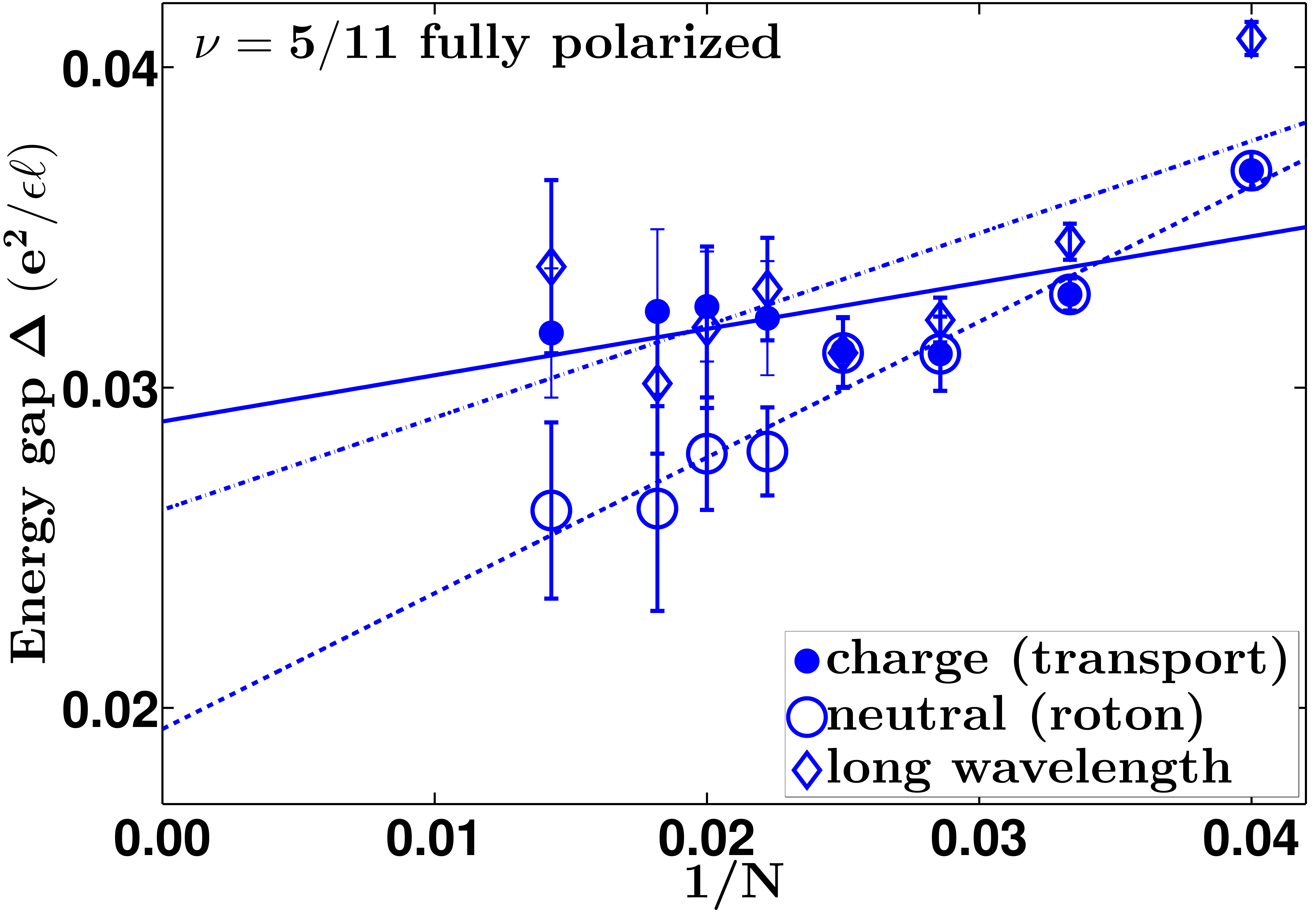}  
\end{center}
\caption{Thermodynamic extrapolation of the charge and neutral LLL Coulomb excitation gaps for fully spin polarized states at various filling factors along the sequence $s/(2s+1)$. For completeness we have also show the long wavelength limit of the CF exciton gap (diamonds).}
\label{gaps}
\end{figure*}

We follow the approach of Jain and Kamilla \cite{Jain97} and plot the CF exciton dispersion (energy difference between the CF exciton states and the ground state) for different system sizes at a given filling factor on the same plot. Fig.~\ref{CF_exciton_dispersion} shows these dispersion curves for different filling factors along the sequence $s/(2s+1)$ from $s=1$ to $s=5$. Fig.~\ref{CF_exciton_dispersion} illustrates that we need to go to systems larger than those accessible to exact diagonalization to see the collapse of the CF exciton dispersion. As Fig.~\ref{CF_exciton_dispersion} and Table.~\ref{tab_minima_positions magnetoroton} illustrate, the positions of the magneto-roton minima agree reasonably well with the predictions from Eqs.~\ref{Simon_Halperin_pred_value} and \ref{Golkar_pred_value} for nearly all filling factors. At small values of $s$ for the strongest minima occuring at the smallest value of $k\ell$, we find that Golkar \emph{et al.}'s prediction agree better with the microscopic CF theory in comparison to Simon and Halperin's \cite{Simon93} numbers, while the subsequent weaker minima agree better with Simon and Halperin's predictions. For small values of $s$, i.e $s=1,2,3$ we find only one magneto-roton minimum in the range $k\ell \lesssim 1$. With increasing value of $s$, there are more and more magneto-roton minima appearing in the range of $k\ell<1$ and they match more accurately with the prediction of Eqs.~\ref{Simon_Halperin_pred_value} and \ref{Golkar_pred_value}, which, as emphasized earlier suggests that these theories work better and better as $s \to \infty$. When $k\ell\rightarrow \infty$, the rotons gradually disappear with decreasing depths, and the dispersion curve flattens out and approaches a constant which gives us the charge gap of the FQH state \cite{Jain07}. The charge, neutral and the long wavelength limit of the CF exciton gaps are shown in Fig.~\ref{gaps} and tabulated in Table~\ref{tab_energy_gaps}. These are in good agreement with previous results \cite{Jain97,Morf02,Park00}. Our estimates for $\nu=4/9$ and $5/11$ are not very reliable as evinced from the large error bars in the individual system results at these filling factors. The positions $k\ell$ of the first and second magneto-roton minima are tabulated in Table.~\ref{tab_minima_positions magnetoroton}. \\

\begin{table*}
\begin{tabular}{|c|c|c|c|c|c|c|c|}
\hline
\multicolumn{1}{|c|}{$\nu$} & \multicolumn{1}{|c|}{$N_{\rm largest}$} & \multicolumn{3}{|c|}{first minima position (in units of $k\ell$)} & \multicolumn{3}{|c|}{second minima position (in units of $k\ell$)} \\ \hline
      &	& microscopic CF & Golkar \emph{et al} [Ref. \cite{Golkar16}] & Simon and Halperin [Ref. \cite{Simon93}] & microscopic CF & Golkar \emph{et al} [Ref. \cite{Golkar16}] & Simon and Halperin [Ref. \cite{Simon93}] \\ \hline
$1/3$ 	& 50	& 1.40	& 1.28	& 1.60	& -	& 2.33	& 2.89	\\ \hline
$2/5$	& 48	& 0.79	& 0.77	& 0.88	& 1.58	& 1.40	& 1.58	\\ \hline
$3/7$	& 60	& 0.49	& 0.55	& 0.61 	& 1.10	& 1.00	& 1.09	\\ \hline
$4/9$	& 64	& 0.36	& 0.43	& 0.46 	& 0.84	& 0.78	&0.83	\\ \hline
$5/11$	& 70	& 0.35	& 0.35	& 0.37	& 0.70	& 0.64	&0.67	\\ \hline
\end{tabular}
\caption{\label{tab_minima_positions magnetoroton}Positions of the first and second magnetoroton minima obtained from the microscopic composite fermion theory for various filling factors along the sequence $s/(2s+1)$ in the $n=0$ LL. Also shown for comparison are the values predicted by Golkar \emph{et al}\cite{Golkar16} and Simon and Halperin\cite{Simon93}. The composite fermion values are obtained from the largest systems (with $N_{\rm largest}$ electrons) considered in this work. The second minima at $\nu=1/3$ is not very clear for the largest system we considered.}
\end{table*}

\begin{figure*}[t]
\begin{center}
\includegraphics[width=7.5cm,height=5.0cm]{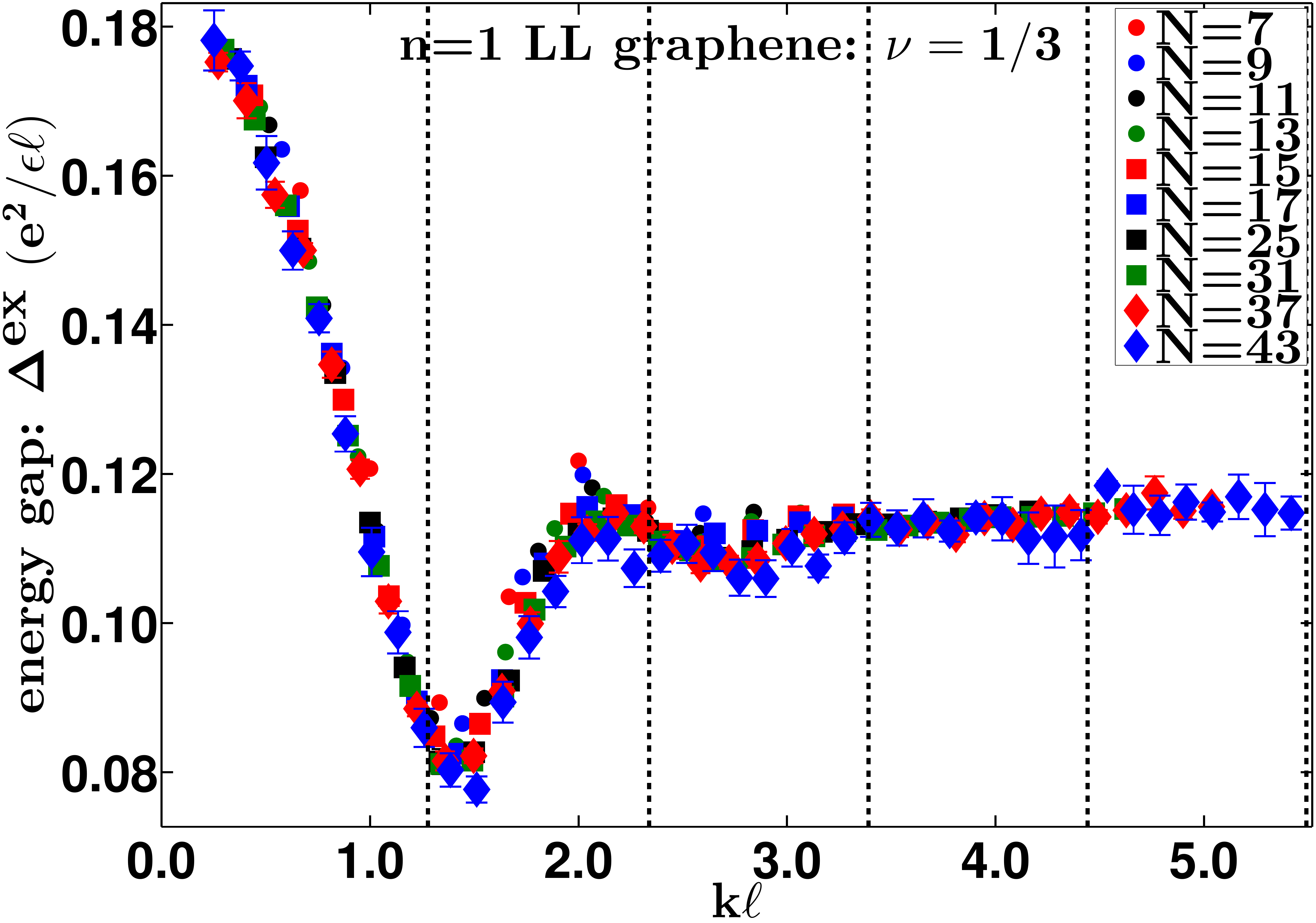}
\includegraphics[width=7.5cm,height=5.0cm]{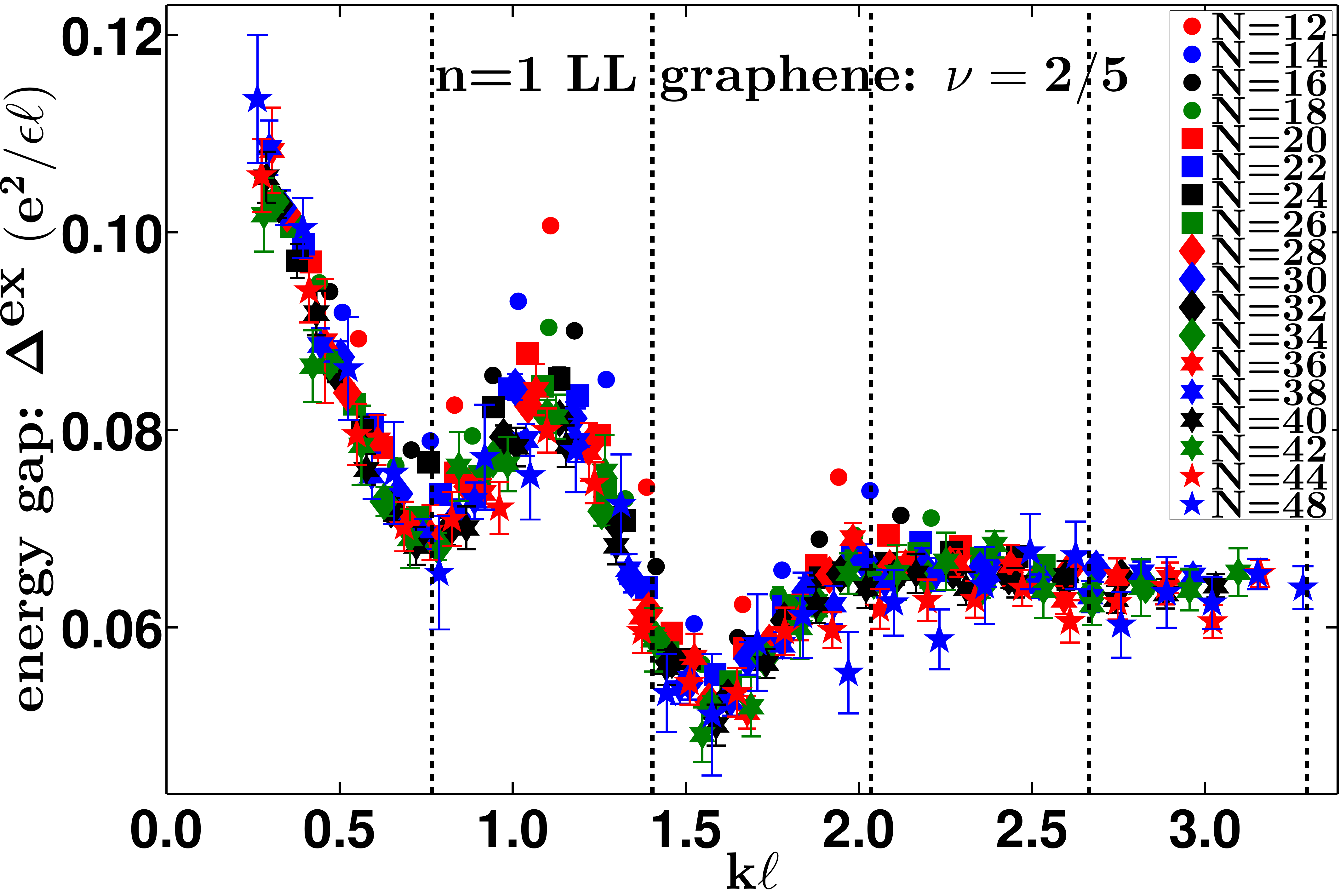} \\
\includegraphics[width=7.5cm,height=5.0cm]{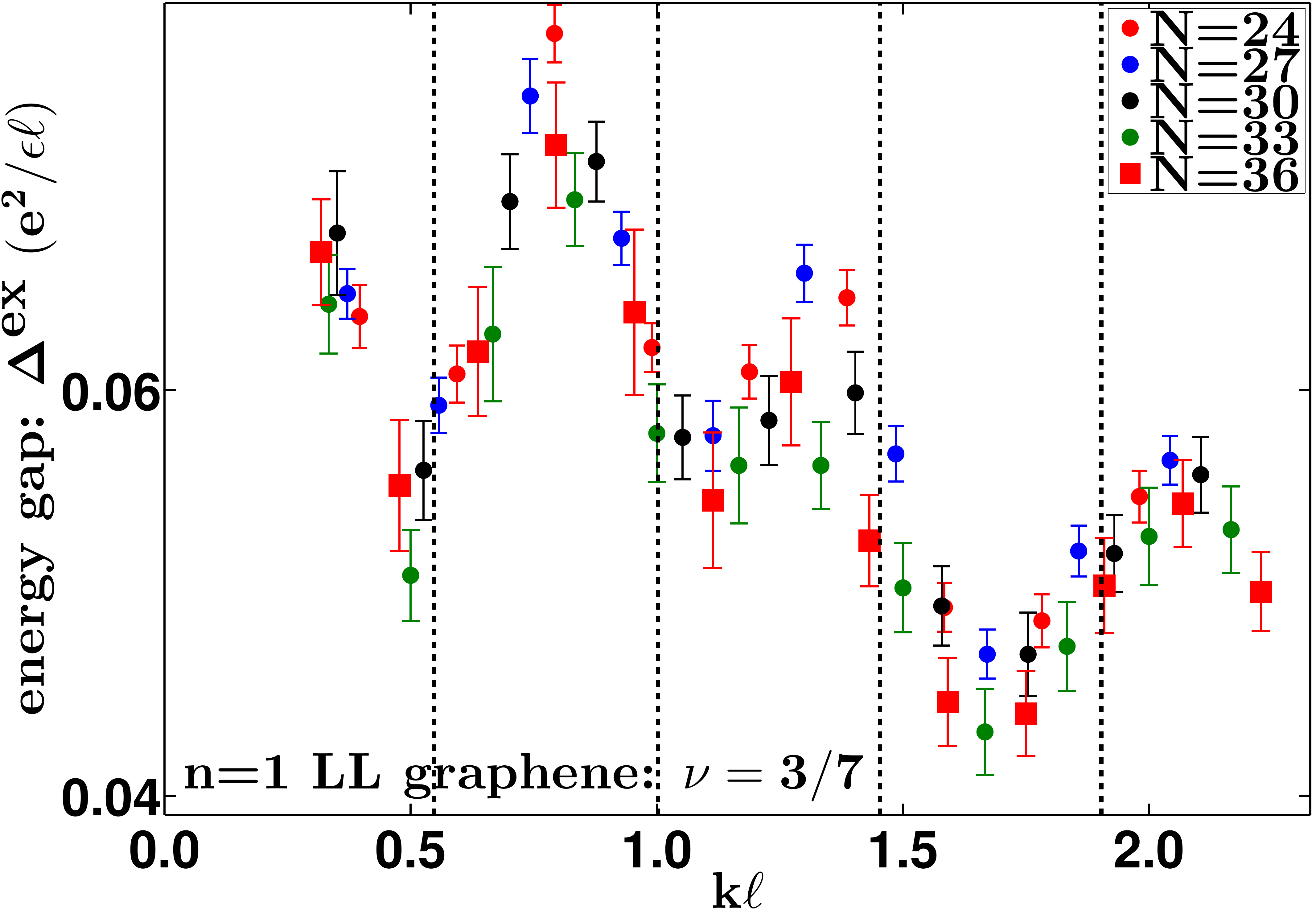} 
\end{center}
\caption{CF exciton dispersion in the $n=1$ LL of graphene for fully spin polarized states at various filling factors along the sequence $s/(2s+1)$ obtained in the spherical geometry. The vertical black dashed lines show the values predicted by Golkar \emph{et al.} \cite{Golkar16}. Some of these exciton dispersions have been reproduced from Ref. \cite{Balram15c}.}
\label{CF_exciton_dispersion_n_1_LL_graphene}
\end{figure*}

\begin{figure*}[t]
\begin{center}
\includegraphics[width=7.5cm,height=5.0cm]{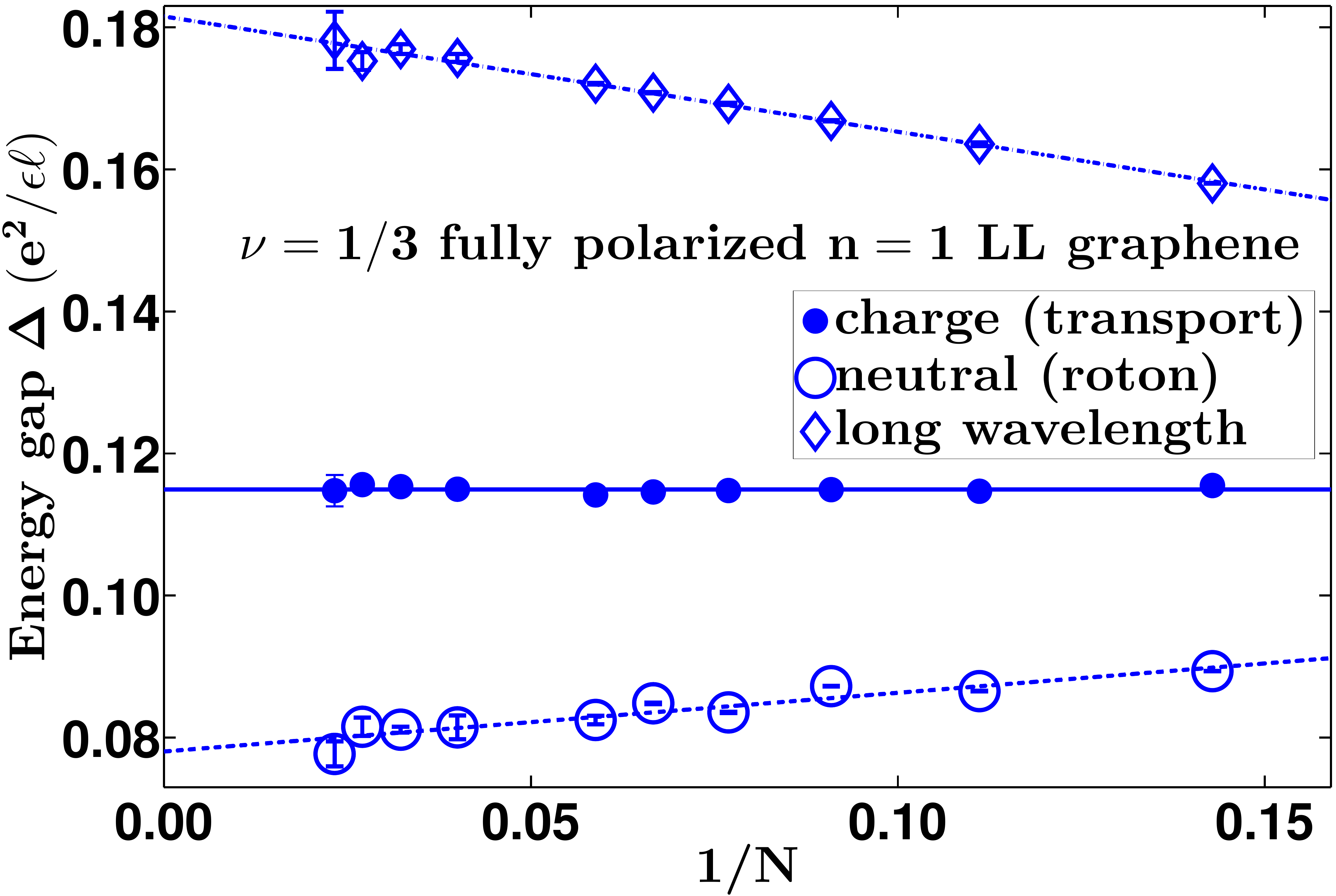}
\includegraphics[width=7.5cm,height=5.0cm]{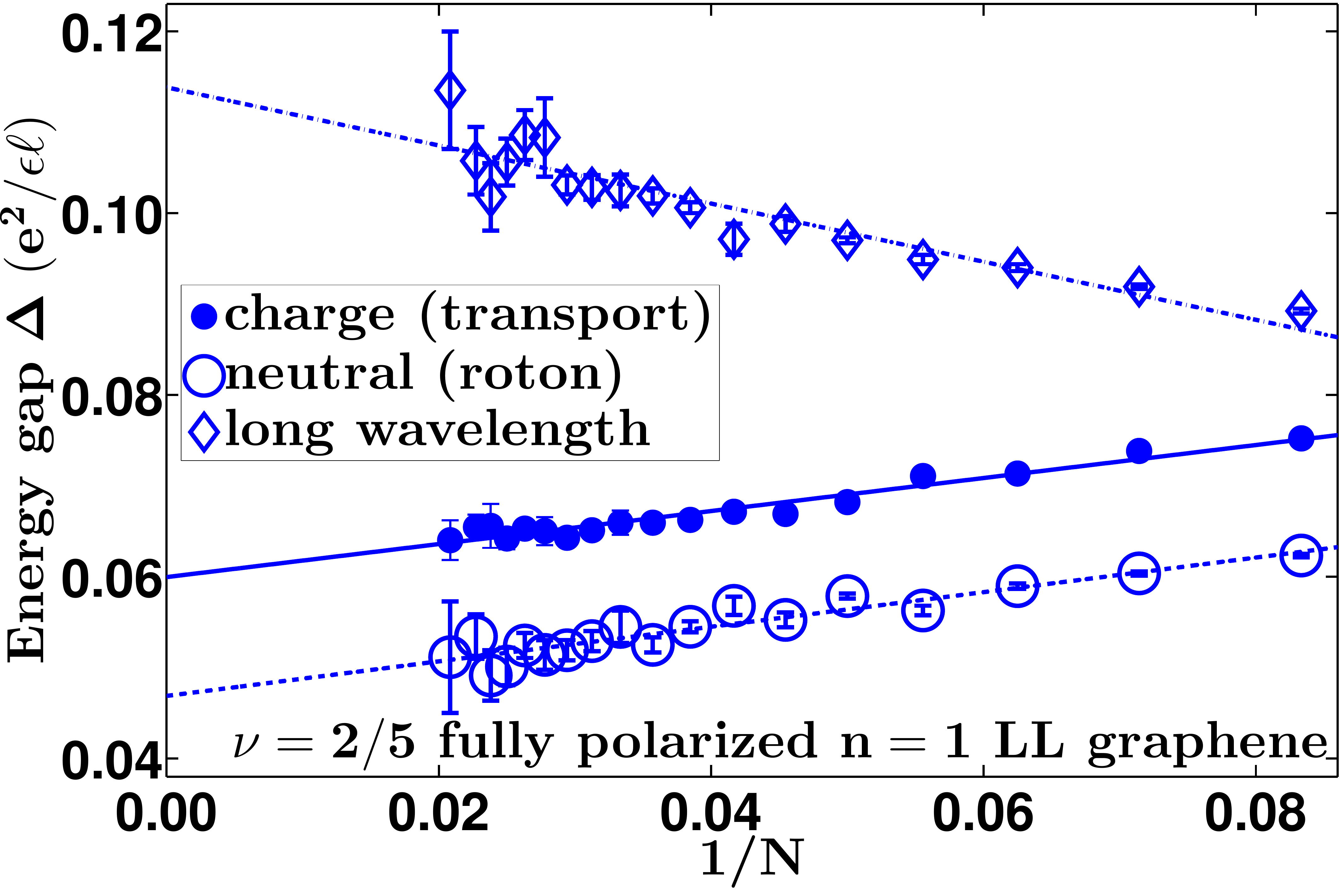} \\ 
\includegraphics[width=7.5cm,height=5.0cm]{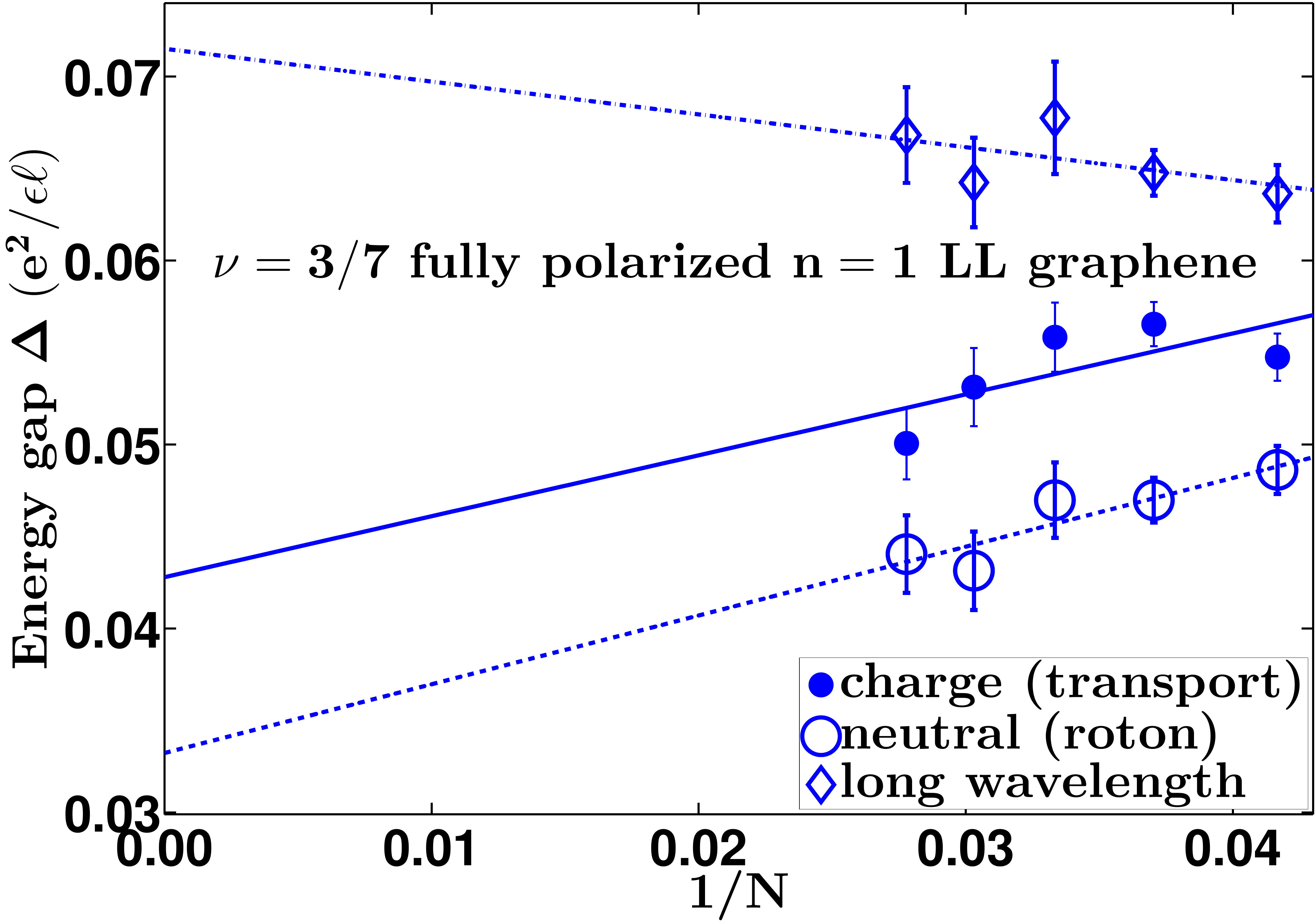}
\end{center}
\caption{Thermodynamic extrapolation of the charge and neutral excitation gaps for fully spin polarized states at various filling factors along the sequence $s/(2s+1)$ in the $n=1$ LL of graphene. For completeness we have also show the long wavelength limit of the CF exciton gap (diamonds).}
\label{gaps_n_1_LL_graphene}
\end{figure*}

We have also calculated the neutral mode dispersions in the $n=1$ LL of graphene up to $s=3$ using the effective interaction of Ref. \cite{Balram15c}. These results are shown in Fig.~\ref{CF_exciton_dispersion_n_1_LL_graphene}. Compared with Fig.~\ref{CF_exciton_dispersion}, it is clear that there is essentially no difference in the positions of magneto-roton minima between the two cases, although they arise from completely different interaction potentials. This supports Golkar \emph{et al.}'s \cite{Golkar16} assertion that the positions of magneto-roton minima do not depend sensitively on the microscopic details of the Halmiltonian. Shibata and Nomura \cite{Shibata09} evaluated the charge gaps for $\nu=1/3$ and $2/5$ in the $n=1$ LL of graphene from an extrapolation of small system results ($N\leq 14$ for $\nu=1/3$ and $N\leq 17$ for $\nu=2/5$) on the spherical and torus geometries respectively. Using the CF exciton dispersion we evaluate the neutral, charge and the $k\rightarrow 0$ limit of the CF exciton gaps for $\nu=1/3,~2/5$ and $3/7$ in the $n=1$ LL of graphene for much larger systems. These results are shown in Fig.~\ref{gaps_n_1_LL_graphene} and tabulated in Table~\ref{tab_energy_gaps}. The numbers for $1/3$ and $2/5$ are consistent with previous results of Ref. \cite{Shibata09}.  \\

\begin{table*}
\begin{tabular}{|c|c|c|c|c|c|c|}
\hline
\multicolumn{1}{|c|}{$\nu$} & \multicolumn{3}{|c|}{gaps in the $n=0$ Landau level [in units of $e^{2}/(\epsilon\ell)$]} & \multicolumn{3}{|c|}{gaps in the $n=1$ Landau level of graphene [in units of $e^{2}/(\epsilon\ell)$]} \\ \hline
	&$k\ell \rightarrow 0$ & neutral (roton) gap & charge (transport) gap, $k\ell \rightarrow \infty$ & $k\ell \rightarrow 0$ & neutral (roton) gap & charge (transport) gap, $k\ell \rightarrow \infty$ \\ \hline
$1/3$ 	& 0.150(1)	& 0.066(1)	& 0.098(1)	& 0.182(1)	& 0.078(1)	& 0.115(1)	\\ \hline
$2/5$	& 0.092(1)	& 0.040(1)	& 0.050(1)	& 0.114(2)	& 0.047(2)	& 0.060(1)	\\ \hline
$3/7$	& 0.066(1)	& 0.031(1)	& 0.035(1)	& 0.071(5)	& 0.033(4)	& 0.043(7)	\\ \hline
$4/9$	& 0.037(2)	& 0.025(1)	& 0.033(1)	& -		& -		& -		\\ \hline
$5/11$	& 0.026(3)	& 0.019(1)	& 0.029(2)	& -		& -		& -		\\ \hline
\end{tabular}
\caption{\label{tab_energy_gaps}Thermodynamic extrapolated values of the long wavelength ($k\ell \rightarrow 0$), neutral (roton) and charge (transport) [long wavevector $k\ell \rightarrow \infty$] Coulomb gaps for various filling factors along the sequence $s/(2s+1)$ in the $n=0$ LL and $n=1$ LL of graphene obtained using the composite fermion theory. The uncertainity shown in the parenthesis is the error in the intercept obtained by doing a linear fit of the finite size gaps with $1/N$, where $N$ is the number of electrons.}
\end{table*}

\begin{figure*}[t]
\begin{center}
\includegraphics[width=7.5cm,height=5.0cm]{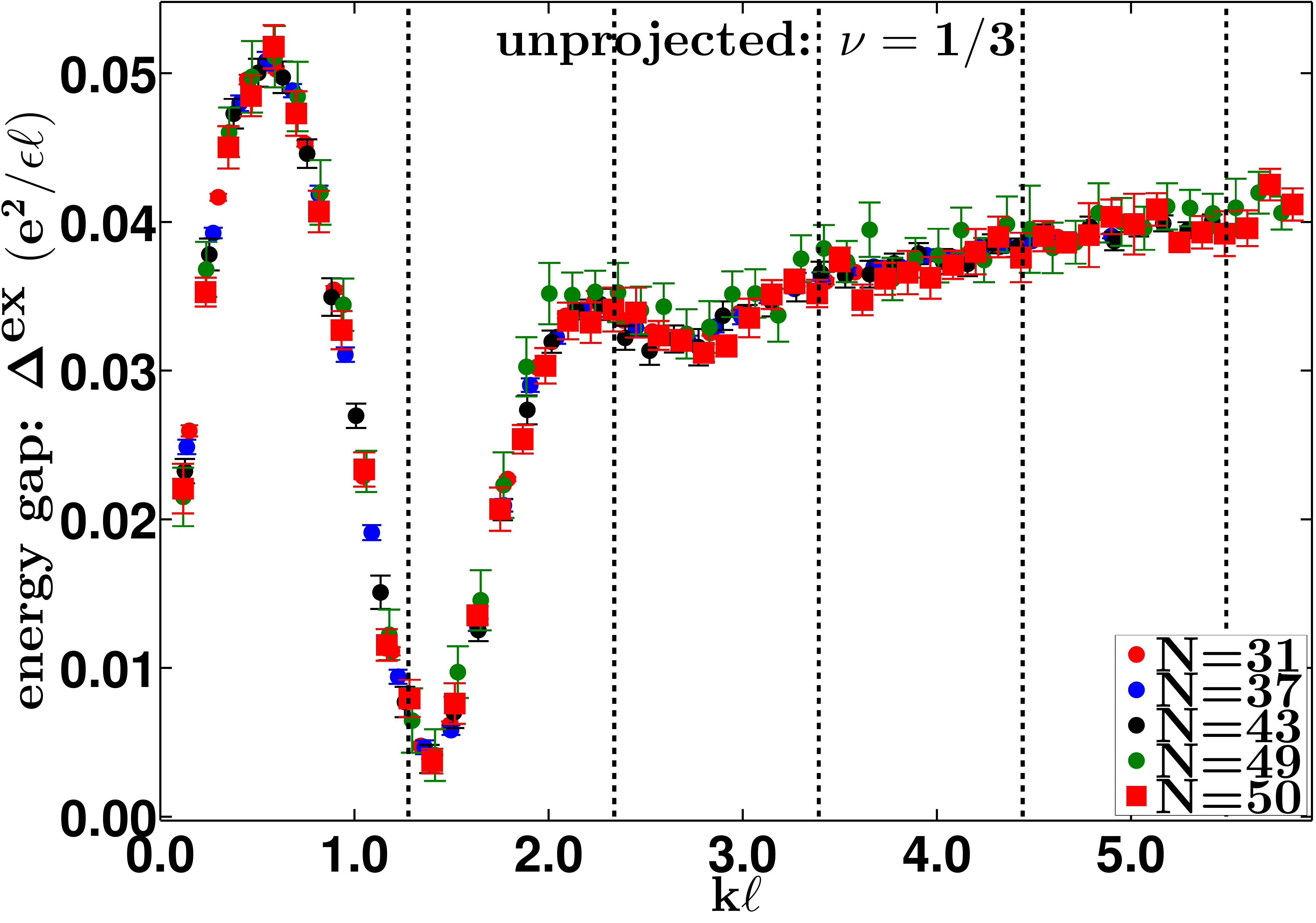} 
\includegraphics[width=7.5cm,height=5.0cm]{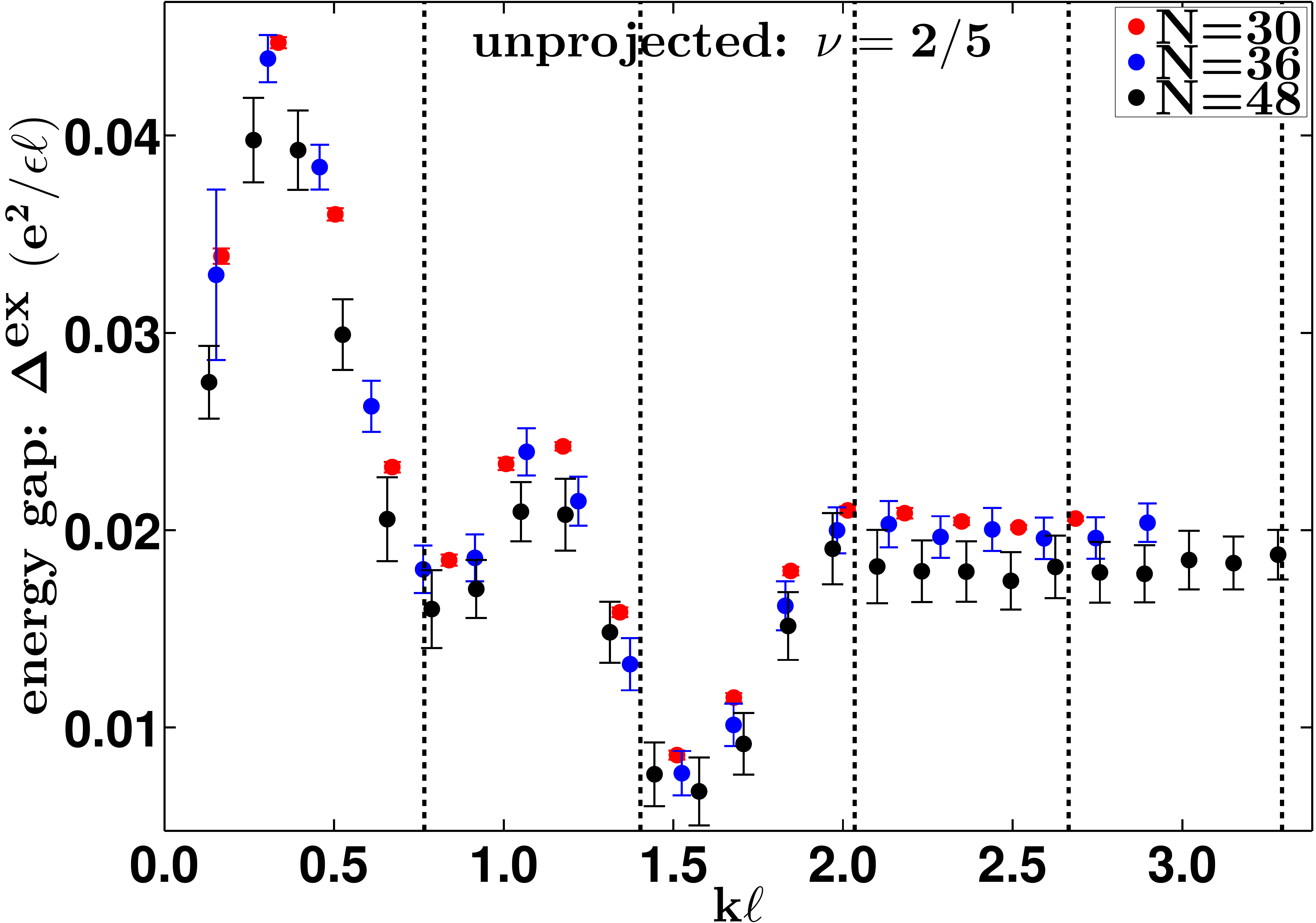} \\
\includegraphics[width=7.5cm,height=5.0cm]{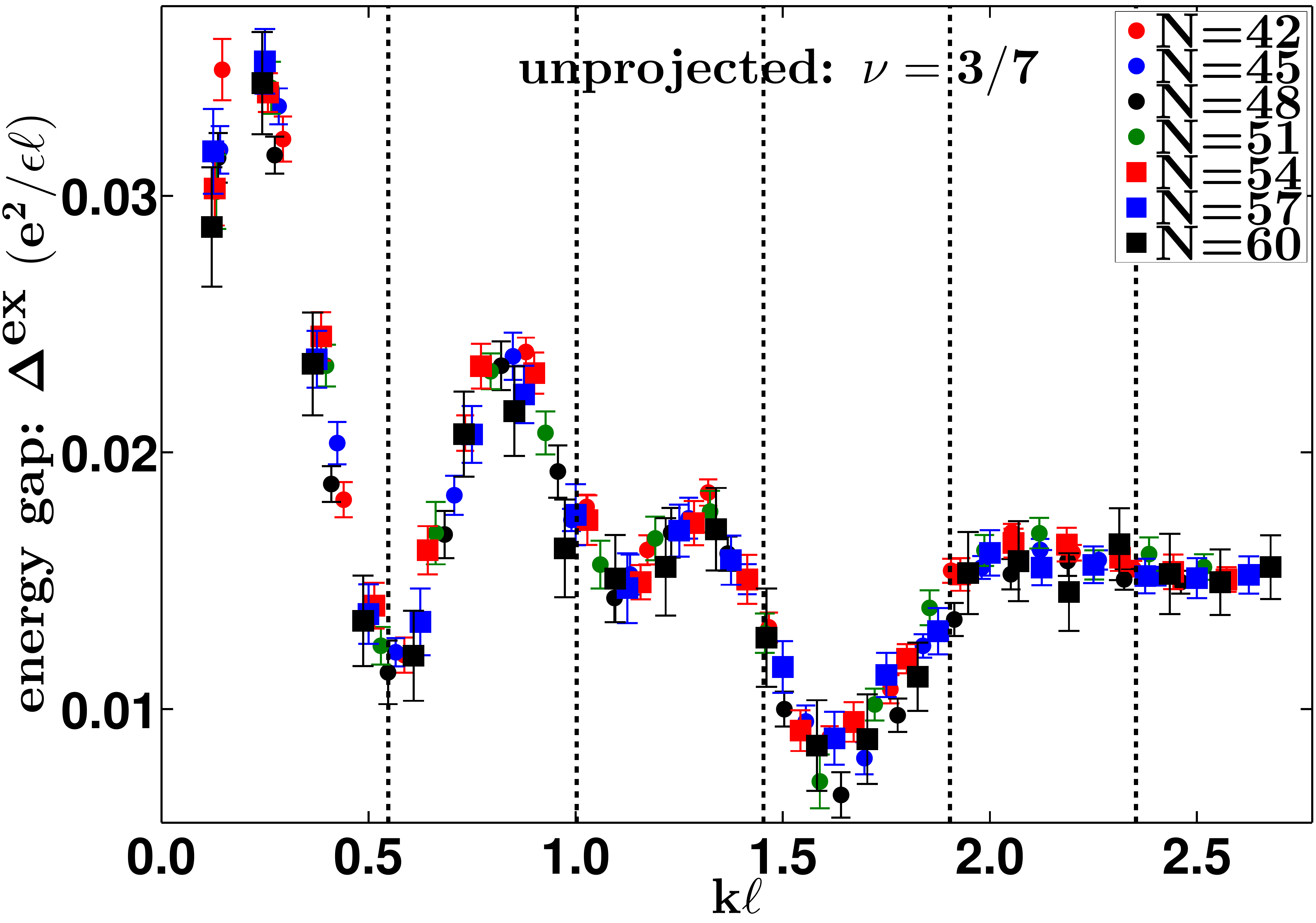} 
\includegraphics[width=7.5cm,height=5.0cm]{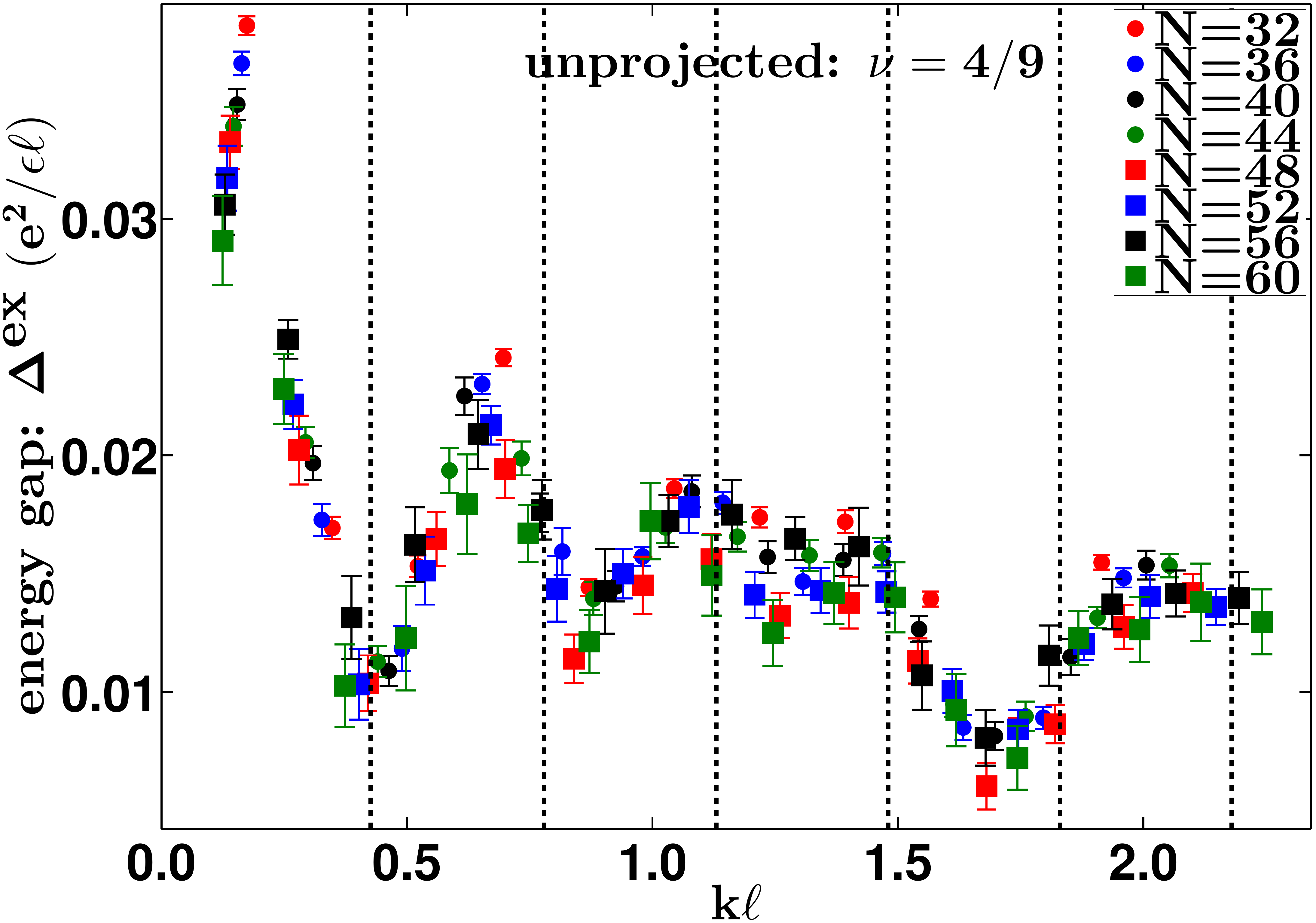}

\end{center}
\caption{CF exciton dispersion for the fully spin polarized unprojected state along the sequence $s/(2s+1)$ obtained in the spherical geometry. The vertical black dashed lines show the values predicted by Golkar \emph{et al.} \cite{Golkar16}. }
\label{CF_exciton_dispersion_unprojected}
\end{figure*}

To test the robustness of the positions of magneto-roton minima to Landau level mixing, we used the unprojected CF wave function to calculate the dispersion of the CF exciton with the Columb interaction at $\nu=1/3,~2/5,~3/7$ and $4/9$. The resultant dispersions are shown in Fig.~\ref{CF_exciton_dispersion_unprojected} and support the assertion that the positions of the magneto-roton minima are not altered by a small admixture with higher LLs. This suggests that the positions of the magnetoroton minima may not be sensitive to small particle-hole symmetry breaking perturbations. \\

\section{Conclusions}
\label{sec:conclusions}
In this work we calculated the dispersion of the the lowest energy spin conserving neutral mode at $\nu=s/(2s+1)$ for $s=1,2,3,4,5$ using the microscopic CF theory, which is known to be quantitatively very accurate. The positions of the magnetoroton minima agree well (within $15\%$) with the recent prediction of these magnetoroton minima positions by Golkar \emph{et al.} \cite{Golkar16}. We also note that the composite fermion Chern-Simons theory \cite{Simon93} predicts the positions of the roton minima that are in good agreement with the microscopic CF theory. We also evaluated the dispersion of the neutral mode for the $n=1$ LL of graphene, where the microscopic CF theory has been shown to work well \cite{Balram15c}. We find that the positions of the magnetoroton minima in the $n=1$ LL of graphene occur at nearly the same values as in the LLL. Furthermore, we evaluated the dispersion of the neutral mode using the unprojected Jain states, which have a small amplitude in higher Landau levels. Here too, we find that the positions of the magnetoroton minima are nearly identical to values obtained in the LLL. Thus our results show that the positions of magneto-roton minima at $\nu=s/(2s+1)$, as predicted by Golkar \emph{et al.}, are independent of the of precise form of the microscopic Hamiltonian.  For completeness, we extended previous works \cite{Jain97,Morf02,Shibata09} by evaluating in detail the charge, neutral and long wavelength gaps of various FQH states in the LLL and the $n=1$ LL of graphene. \\

Golkar \emph{et al.}'s theory can be generalized starting from the $\nu=1/(2p)$ composite fermion Fermi sea to show that the magnetoroton minima at $\nu=s/(2ps+1)$ occur at:
\begin{equation}                                                                                                                                                                                                                                                                                                                                                                 
k\ell=\frac{x_{i}}{2ps+1}
\label{Golkar_extn}
\end{equation}
where $x_{i}$ are the zeros of the Bessel function $J_{1}(x)$. Jain and Kamilla \cite{Jain97} evaluated the CF exciton dispersion in detail at $\nu=1/(2p+1)$ for $p=2,3,4$. The positions of the magnetoroton minima here do not occur close to $k\ell=x_{i}/(2p+1)$. For $\nu=2/13$, we compare the predictions with the CF diagonalization results from Peterson and Jain \cite{Peterson03} for $N\leq 20$. We find that the positions of the first two magneto-roton minima agree quite well with the values predicted by Eq. \ref{Golkar_extn}  but the subsequent ones do not match. As stated above the positions of the magnetoroton minima for $p\geq 2$ could be sensitive to CF $\Lambda$L mixing and therefore more detailed calculations using CF diagonalization for large systems are needed to do a thorough comparison for FQH states described by composite fermions carrying more than two vortices. \\

Finally, we mention here that Golkar \emph{et al.} also make a prediction for the value of the energy scale at the minima in terms of the energy scale at $k=0$. More specifically, they claim that the former is smaller by a power of $n$ than the latter for filling factors along the sequence $n/(2n+1)$. With the filling factors accessible to us, we have not been able to check this prediction accurately. In the future it would be interesting to explore if Golkar \emph{et al.}'s theory can be extended to non-fully spin polarized states, higher energy neutral spin conserving, and spin flip modes for FQH states with various spin polarizations. \\

\emph{Note added: }As this manuscript was nearing completion, we became aware of the recent work of Wang \emph{et al.}\cite{Wang17} who carefully evaluated the locations of the magnetoroton minina using the Chern-Simons theory of composite fermions. They found that the the magnetoroton minima at $\nu=s/(2s+1)$ occur at:
\begin{equation}
k\ell^{\text{CS, Wang \emph{et al.}}}\approx \frac{x_{i}}{2s}\Big(1-\frac{1}{2s}\Big)
\label{Wang_et_al_pred_value_paral_flux}
\end{equation}
and at $\nu=1-s/(2s+1)=(s+1)/(2s+1)$ occur at:
\begin{equation}
k\ell^{\text{CS, Wang \emph{et al.}}}\approx \frac{x_{i}}{2(s+1)}\Big(1+\frac{1}{2s}\Big)
\label{Wang_et_al_pred_value_rever_flux}
\end{equation}
where $x_{i}$ are as before the zeros of the Bessel function of order one of the first kind $J_{1}(x)$. For large values of $s$, the minima positions given by Eq. \ref{Wang_et_al_pred_value_paral_flux} and Eq. \ref{Wang_et_al_pred_value_rever_flux} differ from each other by terms of $\mathcal{O}(s^{-2})$ which shows that the minima positions are particle-hole symmetric to a good accuracy. Furthermore, for large values of $s$ these minima positions approach the value predicted by Eq. \ref{Golkar_pred_value}. 

{\bf Acknowledgment}
We are grateful to J. K. Jain for useful discussions and his comments on the manuscript. The work at Penn State was supported by the U. S. National Science Foundation Grant no. DMR-1401636. ACB thanks the Villum Foundation for support. The Center for Quantum Devices is funded by the Danish National Research Foundation.

{\bf Author contributions}
ACB planned and organized this project; both authors performed theoretical calculations, analysed and discussed the results and contributed to the writing of the manuscript.

\bibliography{../../Latex-Revtex-etc./biblio_fqhe}
\bibliographystyle{apsrev}
\end{document}